\documentclass[%
 reprint,twocolumn,floatfix,
 amsmath,amssymb,
 aps,
prl,
superscriptaddress
]{revtex4-2}

\usepackage{graphicx}
\usepackage{dcolumn}
\usepackage{bm}
\usepackage{color}
\usepackage{xcolor}
\usepackage{comment}
\usepackage[normalem]{ulem}
\usepackage{amsmath}
\usepackage{amssymb}
\usepackage[breaklinks]{hyperref}

\renewcommand{\epsilon}{\varepsilon}

\begin{document}

\title{Active Ionic Fluxes Induce Symmetry Breaking in Charge-Patterned Nanochannels}

\author{Sergi G. Leyva}
\email{sleyva@northwestern.edu}
\affiliation{Center for Computation and Theory of Soft Materials, Northwestern University, Evanston, IL, USA}
\affiliation{Department of Physics and Astronomy, Northwestern University, Evanston, IL, USA}

\author{Ahis Shrestha}
\email{ahis.shrestha@northwestern.edu}
\affiliation{Center for Computation and Theory of Soft Materials, Northwestern University, Evanston, IL, USA}
\affiliation{Department of Physics and Astronomy, Northwestern University, Evanston, IL, USA}

\author{Monica Olvera de la Cruz}
\email{m-olvera@northwestern.edu}
\affiliation{Center for Computation and Theory of Soft Materials, Northwestern University, Evanston, IL, USA}
\affiliation{Department of Physics and Astronomy, Northwestern University, Evanston, IL, USA} 
\affiliation{Department of Materials Science \& Engineering, Northwestern University, Evanston, IL, USA}


\begin{abstract}
Biological systems rely on autonomous modes of charge transport to transmit signals. Instead, conventional synthetic systems typically depend on external fields, such as voltage or pressure gradients, to induce transport, which limits their applicability. Here, we investigate nanochannels in which an electrolyte is confined by
symmetric boundary patterns combining surface charge and active ionic fluxes. We show that the interplay between diffusion, electrostatics, and hydrodynamics in such nano-confined active-charged systems can trigger symmetry breaking above a critical active flux, leading to directed flow. Our results suggest that active-charged nanochannels can generate net flows of the order of hundreds of millimeters per second, opening pathways toward adaptable ionic devices and neuromorphic architectures.
\end{abstract}

\maketitle

Nanofluidic systems hold great promise for enabling novel functionalities 
while significantly improving energy efficiency, from water purification 
\cite{water_filtration} to energy storage \cite{rottenberg,trizac} and 
emulation of synaptic behavior in neuromorphic systems 
\cite{long_term_lyderic,synaptic_real}. These platforms rely on precise 
control of ionic transport at the nanoscale, drawing inspiration from the 
brain, where signal transmission and memory are mediated by 
history-dependent ionic conductivities \cite{synaptic_real}. Most 
nanofluidic studies have focused on passive transport driven by external 
fields \cite{Stein2004,Daiguji2004,vanderHeyden2007,Vlassiouk2007,Siwy2004,
Bocquet2010,Liu2024,Robin2023}, leaving open the question of whether active 
ionic fluxes and electrohydrodynamic couplings can generate emergent 
self-organized transport at the nanoscale. {As a point of reference, when open during the rising phase of an action potential,
a Na$^+$ channel in an axon membrane can conduct $\sim\!10^7$ ions per second
\cite{Sigworth1980} through an effective pore area of order $1$~nm$^2$
\cite{Payandeh2011}. These values imply instantaneous flux densities as large as
$\sim\!100$~mM$\cdot$m/s over timescales of order $1$~ms. Fluxes of 
this magnitude, driven by steep concentration and electrostatic gradients, 
can strongly perturb the local ionic environment, suggesting a possible route toward neuromorphic ionic architectures.}

Recently, it has been proposed that active-charge patterned microchannels 
can generate both ionic and solvent flows \cite{self_ahis}. The activity 
arises solely from the injection and absorption of ions at the channel 
boundaries, a simplified framework reminiscent of that of neuronal axons 
that transmit signals by pumping ions across their membranes 
\cite{HodgkinHuxley1952}. Both the ionic flux and the charge patterns are 
represented as sinusoidal modulations along the channel boundary, with a 
flux--charge phase difference that determines the unidirectional and 
circulatory character of the bulk solvent flow. For symmetric configurations 
with zero flux--charge phase difference, the in-phase pattern symmetry 
cancels the net unidirectional flow.

{As the system enters nanometric confinement, electro-osmotic advection 
can significantly distort the ionic distribution, competing with diffusion 
and electrostatic forces. This confinement also amplifies hydrodynamic 
couplings as the channel width approaches the slip length of the confining 
material \cite{kavokine_2021,lyderic_slip}, producing nonlinear responses, 
including gating, even in flat charge-patterned nanochannels \cite{my_paper}.} {While advective transport can drive symmetry breaking in active pores and
phoretic particles~\cite{malgaretti,michelin_ssb}, its consequences for
ionically active, charge-patterned nanochannels remain largely unexplored.}

\begin{figure*}[t!]
\begin{center}
\includegraphics[width=0.95\textwidth]{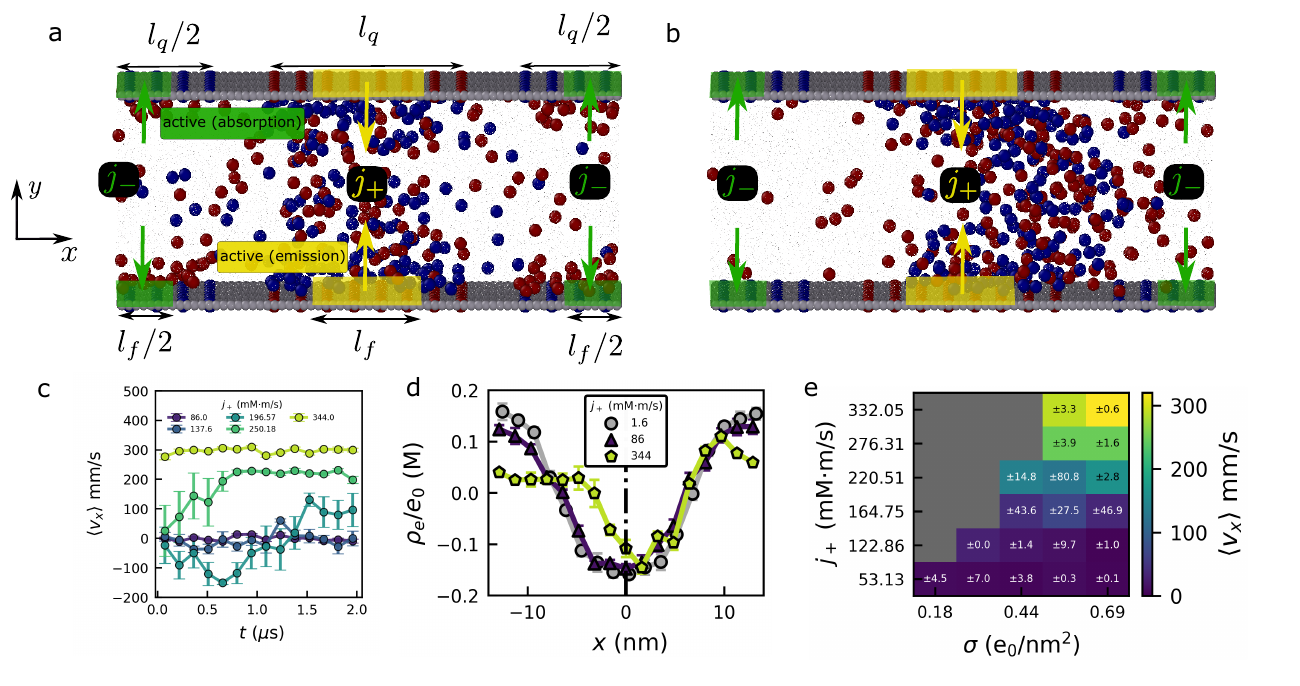}
\end{center}
\caption{DPDS simulations of active ionic fluxes in symmetric charged patterns. All panels correspond to a bulk salt concen-
tration $\rho_0$ = 100 mM. (a) Simulation snapshot showing the {channel geometry, which spans in $x\in[-L/2,L/2]$ and $y\in[-w/2,w/2]$. Grey particles form the walls, the small dots represent the DPD solvent particles, and red and blue particles are cations and anions, respectively. Embedded wall charges impose $\sigma_\pm=\pm0.6\,e_0/\mathrm{nm}^2$. Yellow and green regions denote cation emission and absorption, aligned with the charge pattern. The origin $x=0$ is at the center of the positive patch, while the negative patch is centered at the periodic boundaries $x=\pm L/2$. The active flux is $j_+=86$~mM$\cdot$m/s.}
(b) Snapshot above onset, showing an asymmetric ion distribution and net flow for $j_+=196$~mM$\cdot$m/s.
(c) Time evolution of the average solvent velocity $\langle v_x\rangle$ for different active fluxes at $\sigma_\pm=\pm0.6\,e_0/\mathrm{nm}^2$.
(d) Charge density profile along $x$, showing the asymmetric distribution associated with directed flow.
(e) Phase diagram in the $(\sigma,j_+)$ plane, colored by the net velocity. Numbers indicate the standard deviation of the mean, which increases near the transition. Grey regions correspond to fluxes for which the absorption region becomes
depleted of cations, preventing the insertion--removal protocol from maintaining
electroneutrality.
\label{fig:fig1}}
\end{figure*}
Here, we characterize a charge-patterned nanochannel with active boundaries that emit and absorb cations. We focus on symmetric configurations, in which the charge pattern and the active boundary patches are spatially aligned.
We consider a symmetric aqueous electrolyte at room temperature,
$T=293~{\rm K}$, with diffusivities $D_\pm\simeq 10^{-9}~{\rm m^2/s}$,
relative permittivity $\varepsilon_r=80$, dielectric permittivity
$\varepsilon_w=\varepsilon_r\varepsilon_0$, solvent viscosity
$\eta=1~{\rm mPa\,s}$, and a typical slip length $\ell_s=20~{\rm nm}$.
For the bulk ion concentration $\rho_0=100~{\rm mM}$, the Debye length is
$\lambda_D=(\varepsilon_w k_BT/e_0^2\rho_0)^{1/2}\simeq 1~{\rm nm}$.
The electrolyte is embedded
in a nanochannel of length \(L = 25.9\,\mathrm{nm}\) and width \(w = 10.32\,\mathrm{nm}\) with an alternating surface charge pattern of positively and negatively charged patches, each of length \(l_q\), giving a total charged area fraction with respect to the channel area of \(f_q = 2l_q/L\). 
An active cationic flux pattern, spatially aligned with the charge pattern, alternates ion injection and absorption over regions of length \(l_f\), corresponding to an active fraction \(f_c = 2l_f/L\). Unless otherwise stated, we set \(f_q = 0.8\) and \(f_c = 0.5\). {We place the center of the positively charged patch at $x=0$, matching the center of the injection region, while the centers of the alternating negative patches and absorption regions lie at $x=\pm L/2$, where boundary conditions are applied.}

To model particle-based dynamics explicitly accounting for electrostatic and hydrodynamic interactions, we use Dissipative Particle Dynamics with Solvent (DPDS)~\cite{curk2024dpd}. 
The active ionic flux is imposed by randomly inserting $N$ cations in the emission region and removing $N$ cations in the absorption region every $k$ timesteps, which defines
$j_+ = N/(A k\,dt)$ maintaining overall electroneutrality. 
The wall--fluid interactions are chosen to reproduce the desired slip length. Simulation details are in the Supplementary Materials (SM)~\cite{sup_mat}. 

The described geometry for the DPDS simulations is depicted in Fig.~\ref{fig:fig1}a, which shows a snapshot of a simulation with a cation flux $j_+ = 86$ mM$\cdot$m/s. In this case, no net flow is observed. The charge distribution along the channel remains symmetric with respect to the center of the positive patch and close to equilibrium. 
{The balance between advection, diffusion, and electrostatics generates local recirculating flows that prevent unbounded charge accumulation near the injection region and lead to a stationary charge-density profile. }
{Beyond a critical flux, small perturbations in the initially symmetric charge distribution couple nonlinearly to the electro-osmotic flow, giving rise to a finite net solvent velocity and a lateral displacement of the ionic cloud toward one side of the channel, as illustrated in Fig.~\ref{fig:fig1}b. The temporal evolution of the spatially averaged flow velocity $\langle v_x \rangle$  is shown in Fig.~\ref{fig:fig1}c. The figure highlights flows along the positive x direction, but simulations with different initial conditions show no preferred direction (see SM \cite{sup_mat}).  }
\begin{figure*}[t!]
\includegraphics[width=0.95\textwidth]{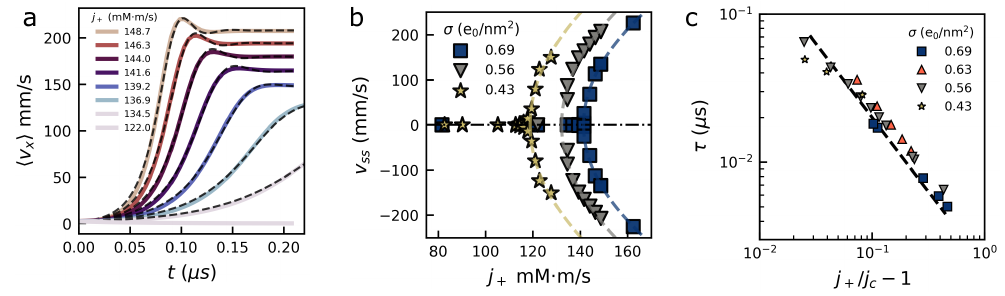}
\caption{  
{(a) Time evolution of the spatially averaged solvent velocity $\langle v_x \rangle$ for increasing cation injection flux $j_+$, obtained for a surface charge density $\sigma_0 = 0.6 e_0/\mathrm{nm}^2$ and bulk salt concentration $\rho_0 = 100$ $\mathrm{mM}$. The dashed lines correspond to fits to Eq.~\ref{eq:landau}. 
(b) Steady-state average velocity $v_{ss}$ as a function of the imposed ionic flux $j_+$.  Above a threshold value $j_c(\sigma_0)$, the system exhibits two symmetry-related branches corresponding to stable flows of opposite direction. The dashed lines are fits to $v_{ss}=c_0\sqrt{j_+-j_c}$. The fitted parameters ($c_0,j_c$) are (49.11 ,141.4), (52.1,132) and (53.92,118) for $\sigma_0$=0.69, 0.56 and 0.44, respectively.
(c) Relaxation time toward the steady state velocity extracted from  fitting the time dependent velocity profiles to Eq.~\ref{eq:landau} as $\tau=1/\lambda$. The dashed line is a fit to a function  $\tau_0/((j_+/j_c)-1)$, where $\tau_0=0.002$ $\mu$s. }  	\label{fig:fig2}}
\end{figure*}

{The corresponding charge distributions, obtained as cross-sectional averages of the net charge density $\rho_e/e_0$ along the channel axis (Fig.~\ref{fig:fig1}d), exhibit a pronounced asymmetry under strong activity. The resulting phase diagram in Fig.~\ref{fig:fig1}e summarizes the onset of spontaneous flow as a function of $\sigma_0$ and $j_+$. Spontaneous flows occur only when both are sufficiently large. Near the transition, long-lived transient flows and occasional reversals produce large run-to-run variability in the time-averaged velocity, reflected in the large standard deviation in Fig.~\ref{fig:fig1}e. Consistently, the
autocorrelation time of the velocity fluctuations peaks in this region,
reflecting slow relaxation near the onset of symmetry breaking (see
SM~\cite{sup_mat}). Although we focus here on fixed surfaces with equal and opposite charge densities, additional simulations show that finite net flows persist for unequal magnitudes of the positive and negative surface charges, as well as when the wall charge is allowed to fluctuate through local charge-regulation dynamics (see SM~\cite{sup_mat}). This supports that the symmetry-breaking mechanism is robust to moderate variations in the surface charging model.}  

{The observed phenomenology bears the signature of a spontaneous
symmetry breaking mediated by a competition between advection,
diffusion and electrostatic interactions. To understand the onset of this
symmetry breaking, we employ a continuum model based on the
Lattice--Boltzmann (LB) method for hydrodynamics, coupled to
electrostatics via a Poisson solver~\cite{ludwig} and to ion transport
through the Nernst--Planck equations~\cite{capuani2004}. This mean-field framework is appropriate for our parameters, since the
electrostatic coupling parameter
$\Xi=2\pi\sigma_0 l_B^2\simeq1.9$, where
$l_B=e_0^2/(4\pi\varepsilon_w k_BT)$ is the Bjerrum length, is well below the
strong-coupling threshold $\Xi\simeq10$~\cite{Moreira_2000}, consistent with the applicability of mean-field approximations.} The LB model is parameterized to match the physical units of the DPDS system and is coupled self-consistently to the Poisson and Nernst--Planck equations to describe electrostatic and ionic transport. We use the same channel geometry, charge pattern, and active regions as in the DPDS simulations, corresponding to Fig.~\ref{fig:fig1}a. Active fluxes are imposed by adding and removing cation density at the emission and absorption boundaries, respectively, yielding a steady flux $j_+$ (see SM for details ~\cite{sup_mat}). The initial condition consists of slightly different uniform densities on the left and right halves of the channel, $\rho_l=\rho_0+0.025\rho_0$ and $\rho_r=\rho_0-0.025\rho_0$, respectively.

\begin{figure*}[ht!]
\begin{center}
\includegraphics[width=\textwidth]{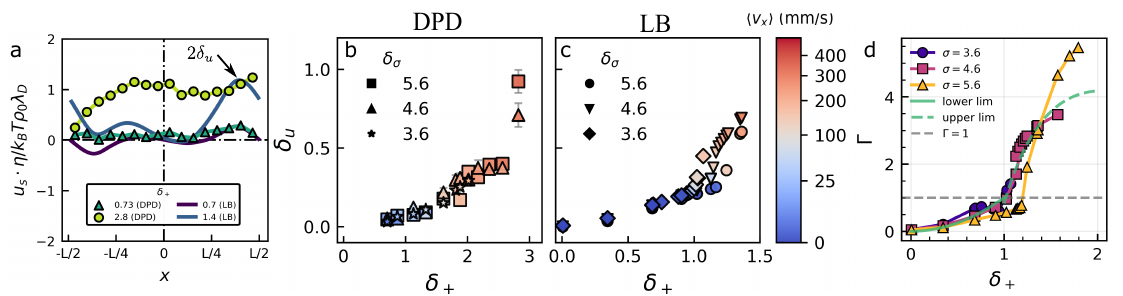}
\end{center}
\caption{{(a)~Dimensionless slip velocity profile along the channel from LB
(lines) and DPD (symbols) simulations, for representative values of
$\delta_+$ below, near, and above the transition. The arrow indicates the
extracted amplitude $\delta_u$ in LB, while in DPD it corresponds to the maximum slip velocity. (b,\,c)~Dimensionless slip amplitude
$\delta_u$ as a function of the flux amplitude $\delta_+$ for different
surface charges $\delta_\sigma$, from (b)~DPD and (c)~LB simulations.
The color scale indicates the steady-state velocity. (d)~Transport
ratio $\Gamma$ as a function of $\delta_+$ from LB simulations. Solid
and dashed green lines show the analytical predictions from the
low-flux ($\Gamma \sim \delta_+^2$) and near-critical (Eq.~\ref{eq:gamma}) regimes,
respectively. The symmetry breaking is observed only for
$\Gamma > 1$.}
\label{fig:fig3}}
\end{figure*}

As shown in Fig.~\ref{fig:fig2}a, the simulations reveal the onset of symmetry breaking as the activity increases, producing a net flow whose direction depends on the initial density perturbation. {The time-dependence of the net flow velocity is well captured by the normal form of a
supercritical pitchfork bifurcation \cite{Strogatz1994}, with respect to the $\langle v_x \rangle \to -\langle v_x \rangle$ symmetry,}
\begin{equation}
\varepsilon \ddot{\langle v_x \rangle} + \dot{\langle v_x \rangle} = a \langle v_x \rangle + b \langle v_x \rangle^3 .
\label{eq:landau}
\end{equation}
{Here, $a$ characterizes the competition between advective transport and the restoring effects of diffusion and electromigration.
The parameter $b$ describes the nonlinear saturation arising from the displacement of the ionic cloud: as the flow grows, the electrostatic restoring force eventually balances the advective driving. Finally, $\varepsilon$ is a timescale associated with the advective displacement of the ionic cloud before it relaxes towards the steady state, resulting in a small overshoot.}
{The parameters $\varepsilon$, $a$, and $b$ are obtained by fitting
Eq.~\eqref{eq:landau} to the time-dependent velocities in
Fig.~\ref{fig:fig2}a (see SM~\cite{sup_mat}). A transition to a finite flow occurs for $a > 0$. The steady-state
solution $v_{ss} = \sqrt{-a/b}$, combined with the generic linear
vanishing of $a$ near the bifurcation, $a \propto j_+ - j_c$
(confirmed by the fits, see SM~\cite{sup_mat}), yields the
mean-field scaling $v_{ss} \propto \sqrt{j_+ - j_c}$, consistent with Fig.~\ref{fig:fig2}b.
Linearising Eq.~\eqref{eq:landau} around $\langle v_x \rangle = 0$ yields
the characteristic equation $\varepsilon \lambda^2 + \lambda - a = 0$,
with roots $\lambda_\pm = (-1 \pm \sqrt{1+4\varepsilon a})/(2\varepsilon)$.
After a brief transient on the timescale $\sim 2\varepsilon$ during which
the stable mode $\lambda_-$ decays, the early-time dynamics is dominated
by the unstable mode, $\langle v_x \rangle(t) \propto e^{\lambda_+ t}$.
Near the transition ($\varepsilon a \ll 1$), $\lambda_+ \approx a$, and
the characteristic growth time $\tau = 1/\lambda_+ \approx 1/a$ diverges
as $j_+ \to j_c$, revealing critical slowing down near the bifurcation. This behaviour is consistent with the DPD simulations, in
which intermittent flow reversals produce long deviations from the mean
and increased autocorrelation times of velocity fluctuations.}
{The onset of symmetry breaking in LB simulations appears at slightly smaller
fluxes than in DPD simulations, likely because thermal fluctuations
broaden the transition and perturb the developing charge asymmetry. For similar simulation parameters both methods yield consistent charge
distributions and flow patterns (see SM~\cite{sup_mat}).}

{We identify the onset mechanism as a positive feedback loop,
which follows the sequence (i) $\rightarrow$ (ii) $\rightleftarrows$
(iii) $\cdots$ $\rightarrow$ (iv): (i) a small perturbation in the salt distribution generates a lateral electro-osmotic flow via the body force; (ii) this flow advects the ionic cloud laterally, amplifying the charge asymmetry through advective coupling; (iii) the amplified charge asymmetry drives a stronger flow, reinforcing step (ii), until the feedback saturates; and (iv) saturation occurs when the distortion of the charged cloud is balanced by electro-diffusive relaxation \cite{sup_mat}. 
To further understand the consequence of this advective feedback cycle, we extend the analytical framework of Ref.~\cite{self_ahis} coupling the Poisson, Nernst--Planck, and Stokes equations to include advective effects and slip velocity conditions (see SM~\cite{sup_mat} for the detailed
derivation and comparison with simulations). The surface charge and 
active cationic flux are modeled as in-phase cosine modulations,
$\sigma(x) \approx \sigma_0\cos(kx)$ and $j_+(x) \approx j_{0+}\cos(kx)$, respectively, with
wavenumber $k \approx 2\pi/L$. Two dimensionless control
parameters emerge from the imposed boundary conditions: the flux amplitude
$\delta_+ = j_{0+}\lambda_D/(D\rho_0)$ and the charge amplitude
$\delta_\sigma = \sigma_0 e_0\lambda_D/(\varepsilon_w k_BT)$.}

{The third relevant scale arises from the system's own response.
Above the bifurcation, the LB simulations develop a non-zero slip velocity
at the channel walls, well described by $u_s(x) \approx u_0[1 - \cos(kx)]$,
with amplitude $u_0$. This emergent slip defines an advective coupling
$\delta_u = u_0\eta/(k_BT\rho_0\lambda_D)$, which characterizes the
strength of the self-generated flow at the boundary. The slip-velocity
profile thus provides the connection between the continuum theory and
the numerical results: Figure~\ref{fig:fig3}a shows $u_s(x)$ extracted from the simulations, made
dimensionless by the natural scale $k_BT\rho_0\lambda_D/\eta$, along
the channel for representative cases in both DPD and LB. For
$\delta_+ \lesssim 1$ or near $\delta_+ \approx 1$, the slip profile is
zero or near-zero, producing only recirculating flows. For
$\delta_+ \gtrsim 1$, a symmetry-breaking slip mode develops, and
$\delta_u$ becomes finite.
By incorporating the symmetry-breaking slip as an effective boundary
condition, the perturbative expansion yields $\langle \rho_e \rangle \sim \rho_{\rm sym}
+ \text{Pe}_u (\delta_\sigma - \delta_+)
\left[
\mathcal{F}\sin(kx) + \mathcal{G}\sin(2kx)
\right]$ as the sum of the symmetric cosine components $\rho_{\rm sym}$ and asymmetric sine contributions at second order,}
{where $\mathcal{F}$ and $\mathcal{G}$ are coefficients depending on
$\lambda_D$, $L$, and $w$~\cite{sup_mat}, and
$\text{Pe}_u = \delta_u \varepsilon_w(k_BT)^2/(e_0^2\eta D) \approx 0.47 \ \delta_u$ which we refer to as the self-electrokinetic P\'{e}clet number. This confirms the relation between the slip velocity and the advected charge density shift observed in both simulation methodologies. 
The amplitude $\delta_u$ is
extracted from the maximum value of the dimensionless
slip at the wall in the simulations (indicated in Fig.~\ref{fig:fig3}a).
Figures~\ref{fig:fig3}b--c show that $\delta_u$ increases with
$\delta_+$ in both DPD and LB, with both methods yielding consistent
values.}

{The analytical model yields an explicit criterion for the onset of
symmetry breaking through the transport ratio $\Gamma$, defined as the ratio
of the advective ionic current $\rho_e v_x$ to the combined
diffusive and electromigration fluxes, averaged across
the channel width (see SM~\cite{sup_mat}). Although the perturbative expansion is formally valid for $\delta_\sigma \ll 1$, the charge profiles below the onset remain close to equilibrium, supporting $\Gamma$ as a qualitative predictor of the transition.}
{Before a net flow develops
($\delta_u \sim 0$), the flow arises only through nonlinear electrokinetic
coupling at second order, giving $\Gamma \sim \delta_+^2$. This
implies that $\Gamma = 1$ is reached when $\delta_+ \sim 1$, i.e., \
when the active flux becomes comparable to the diffusive flux scale
$D\rho_0/\lambda_D$. Once the symmetry is broken and a
finite slip develops, the transport ratio takes the explicit form}
\begin{equation}\label{eq:gamma}
\Gamma \sim
\frac{2\tilde{k}\,\delta_u\,|\delta_+ - \delta_\sigma|}
{(1+\tilde{k}^2)\,\delta_+},
\end{equation}
{where $\tilde{k} = \lambda_D k$. As shown in Fig.~\ref{fig:fig3}d,
both the low-flux scaling and the near-critical expression are
consistent with the simulation data: the transition from $\Gamma < 1$
to $\Gamma > 1$ occurs near $\delta_+ \sim 1$, and
Eq.~\eqref{eq:gamma} provides an upper bound that captures the
behavior across different surface charges.} {This analysis identifies the parameter $a$ in Eq.~\eqref{eq:landau}
with $a \propto \Gamma - 1$: the linear growth rate of the instability
is governed by the excess of advective over electro-diffusive transport.
Combined with the low-flux scaling $\Gamma \sim \delta_+^2$, which gives
$\Gamma - 1 \propto j_{0+} - j_c$ near the transition, this yields
$\tau \sim 1/a \propto 1/(\Gamma - 1) \propto 1/(j_{0+} - j_c)$,
recovering the critical slowing down observed in the LB fits
(Fig.~\ref{fig:fig2}c). The theory further predicts that the
charge asymmetry decays with increasing $w/\lambda_D$ (see
SM~\cite{sup_mat}), suppressing the net flow, a trend confirmed
in both DPD and LB simulations.}

{In summary, we have shown that an electrolyte confined in a nanochannel with symmetric active-charged boundary patterns can undergo a symmetry breaking that generates spontaneous directed flows without any external pressure or voltage gradient, sustained by an advective feedback mechanism. Combining particle-based and continuum simulations with analytical theory, we identify the bifurcation as governed by the competition between advective ion transport and electro-diffusive restoring forces, captured by the transport ratio $\Gamma$
. The transition occurs when the active ionic flux reaches the diffusive flux scale ($\delta_+ \sim 1$
), and the resulting flows can reach velocities on the order of hundreds of millimetres per second. The fixed-charge pattern provides a minimal setting to isolate this feedback. Additional nanoscale interfacial effects, such as dielectric variations near the walls, may reshape this electrostatic coupling in channels narrower than a few nanometers, where the perpendicular permittivity is known to deviate significantly from its bulk value \cite{kavokine_2021}. Yet the instability is not tied to the microscopic surface charging model: simulations with unequal charge magnitudes and local charge-regulating walls show that flows persist when the average surface charge is sufficiently large.}

{Taken together, these results show that autonomous, self-organised ionic transport reminiscent of biological signaling can emerge in synthetic nanoconfined geometries. Unlike voltage-driven nanofluidic memristors~\cite{Kamsma2023,Emmerich2024}, here the net flows are sustained by an active ionic flux, closer in spirit to chemically powered biological ion pumping. More broadly, this suggests that active ionic boundary conditions can generate tunable directed nanoscale transport through self-organized electrokinetic feedback, rather than through externally imposed voltage or
pressure gradients.}

\begin{acknowledgments} 
We thank Paolo Malgaretti for insightful initial discussions. 
This work was supported by the National Science Foundation (NSF) under Grant No.~DMR-2452280.
\end{acknowledgments}

\bibliography{nanochannel,chargeregulation,journals}

\end{document}